\documentclass[conference,9pt]{IEEEtran}

\usepackage[utf8]{inputenc}
\usepackage{booktabs} 
\usepackage{graphicx}
\usepackage{subfigure}
\usepackage{units}
\usepackage{blindtext}
\usepackage{xcolor,xspace}
\usepackage{amsmath}
\usepackage{paralist}
\usepackage{amsmath}
\usepackage{tikz}
\usepackage{booktabs}
\usepackage{pifont}
\usepackage{multirow}
\usepackage{hhline}
\usepackage{balance}
\usepackage{footnote}
\usepackage{logicproof}
\usepackage{amsthm}
\usetikzlibrary{patterns}
\usetikzlibrary{calc}
\usetikzlibrary{decorations.pathreplacing}

\usepackage{arydshln}
\usepackage{mdframed}
\usepackage{pgf-umlsd}

\usepackage[T1]{fontenc}
\usepackage{
  beramono,
  listings,
  textcomp
}

\usetikzlibrary{arrows,automata}
\usetikzlibrary{calc}
\usetikzlibrary{arrows,calc,shapes,decorations.pathreplacing}

\usepackage{listings}
\usepackage[ruled]{algorithm2e}
\usepackage{algorithmic}

\LinesNumbered

\usepackage{url}

\expandafter\def\csname ver@l3regex.sty\endcsname{}
\usepackage[
    n,
    advantage,
    operators,
    sets,
    adversary,
    landau,
    probability,
    notions,
    logic,
    ff,
    mm,
    primitives,
    events,
    complexity,
    asymptotics,
    keys]{cryptocode}

\newcommand{\acro}{{{\sf\it Tiny-CFA}}\xspace}

\newcommand{\vrased}{{{\sf\it VRASED}}\xspace}

\newcommand{\dev}{{\ensuremath{\sf{\mathcal Prv}}}\xspace}
\newcommand{\prv}{{\ensuremath{\sf{\mathcal Prv}}}\xspace}
\newcommand{\vrf}{{\ensuremath{\sf{\mathcal Vrf}}}\xspace}

\newcommand{\RA}{{\ensuremath{\sf{\mathcal RA}}}\xspace}
\newcommand{\CFA}{{\ensuremath{\sf{\mathcal CFA}}}\xspace}

\newcommand{\chal}{{\ensuremath{\sf{\mathcal Chal}}}\xspace}

\newcommand{\attkey}{\ensuremath{\mathcal K}\xspace}

\newcommand{\rtop}{\ensuremath{\mathcal R}\xspace}

\newcommand{\program}{\ensuremath{\mathcal P}\xspace}

\renewcommand\adv{\ensuremath{\sf{\mathcal Adv}}\xspace}

\mathchardef\mhyphen="2D

\newcommand{\cflog}{{\small CF-Log}\xspace}

\begin{document}

\title{\acro: A Minimalistic Approach for Control-Flow Attestation Using Verified Proofs of Execution}

\author{\IEEEauthorblockN{Ivan De Oliveira Nunes}
\IEEEauthorblockA{\textit{University of California, Irvine} \\
ivanoliv@uci.edu}
\and
\IEEEauthorblockN{Sashidhar Jakkamsetti}
\IEEEauthorblockA{\textit{University of California, Irvine} \\
sjakkams@uci.edu}
\and
\IEEEauthorblockN{Gene Tsudik}
\IEEEauthorblockA{\textit{University of California, Irvine} \\
gene.tsudik@uci.edu}
}
\maketitle
%
\begin{abstract}
The design of tiny trust anchors attracted much attention over the past decade, to secure
low-end MCU-s that cannot afford more expensive security mechanisms. In particular, hardware/software (hybrid) 
co-designs offer low hardware cost, while retaining similar security guarantees as (more expensive) 
hardware-based techniques. Hybrid trust anchors support security services (such as remote attestation, proofs of 
software update/erasure/reset, and proofs of remote software execution) in resource-constrained MCU-s, e.g., 
MSP430 and AVR AtMega32.
Despite these advances, detection of control-flow attacks in low-end MCU-s remains a challenge, since hardware 
requirements for the cheapest mitigation techniques are often more expensive than the MCU-s themselves.
In this work, we tackle this challenge by designing \acro~-- a Control-Flow Attestation (\CFA) technique with a 
single hardware requirement -- the ability to generate proofs of remote software execution (PoX).
In turn, PoX can be implemented very efficiently and securely in low-end MCU-s.
Consequently, our design achieves the lowest hardware overhead of any
\CFA technique, while relying on a formally verified PoX as its sole hardware requirement.
With respect to runtime overhead, \acro also achieves better performance than prior \CFA techniques based 
on code instrumentation. We implement and evaluate \acro, analyze its security, and demonstrate its 
practicality using real-world publicly available applications.
\end{abstract}

\section{Introduction}\label{sec:intro}
With the growth of the Internet of Things (IoT) and popularity of Cyber-Physical Systems (CPS), embedded devices 
have become ubiquitous in modern society. Since they often perform safety-critical tasks and process security- and 
privacy-sensitive data, they become an attractive attack targets. In this context, Remote Attestation (\RA) has been 
proposed as a means to secure the software state of embedded systems. \RA is a challenge-response protocol 
(see Section~\ref{sec:background_ra} for details) whereby a trusted verifier (\vrf) obtains an authentic and timely report 
about the software state of an untrusted (and potentially infected) remote device, called prover (\prv). This report allows 
\vrf to learn whether \prv's current state is trustworthy, i.e., whether it hosts benign software.
\RA has been implemented efficiently, even on low-end MCU-s~\cite{smart,Sancus17,vrasedp} to detect malware 
presence in the form of modified executables. However, conventional (aka static) \RA can only ensure 
integrity of binaries and not of their execution.

Runtime/data-oriented attacks~\cite{runtime_attacks_sok} tamper with execution state on the program's stack or 
heap to arbitrarily divert the program's execution flow. Such attacks need not modify the executable itself, but only 
the order in which its instructions are executed. Thus, they are not detectable by conventional \RA. In particular, 
\RA cannot detect runtime software attacks that hijack the program's control-flow. Control-flow attacks can be launched 
by a variety of means. For instance, in languages such as \texttt{C}, \texttt{C++}, and Assembly (which are widely used 
to program MCU-s), buffer overflows~\cite{cowan2000buffer} can overwrite functions' return addresses, hijacking the 
program's control-flow and launching well-known Return-Oriented Programming (ROP) attacks~\cite{rop}. These attacks 
are especially dangerous for low-end MCU-s that can not avail themselves of more sophisticated OS-based mitigations, e.g., 
canaries, Address Space Layout Randomization (ASLR), and Control-Flow Integrity (CFI) techniques, available in 
high-end platforms. We discuss a concrete example of such an attack in low-end MCU-s (and how it is detected by 
\acro) in Section~\ref{sec:case_studies}.

Control-Flow Attestation (\CFA)~\cite{cflat,dessouky2017fat,dessouky2018litehax,zeitouni2017atrium} augments 
conventional \RA capability to enable detection of control-flow attacks. In a nutshell, \CFA techniques provide \vrf with a 
report that allows it not only learn if the expected code is loaded on \prv, but also which particular instruction path was 
taken during each execution of this program. In other words, \CFA provides \vrf with an authentic and unforgeable report 
that allows \vrf to learn if instructions of a given program were executed in a particular expected/legal order, or a set thereof. 
This is typically achieved by securely logging information associated with the destination of each control-flow altering instruction,
e.g., \texttt{jumps}, \texttt{branches}, \texttt{returns}, during program execution.

\CFA techniques have been implemented on medium- to high-end embedded devices (e.g., Raspberry Pi, and RISC-V based processors), 
by leveraging trusted hardware support, such as ARM TrustZone, hardware branch monitors, and hardware hash engines. 
However, for resource constrained MCU-s, these requirements are too costly, since their hardware overhead is often higher than 
that of the MCU's core itself, in terms of size, energy and monetary cost. To bridge this gap, our work leverages a recently proposed 
primitive -- Proofs of Execution -- PoX~\cite{apex} (see Section~\ref{sec:background_pox} for details) -- along with automatic 
code instrumentation, to derive a new \CFA technique. Since PoX can be implemented efficiently even on most resource-constrained 
MCU-s, our \CFA technique has considerably lower hardware overhead than that of prior work.

\textbf{Contribution:} we design, implement, and evaluate \acro -- a \CFA technique based on automated software 
instrumentation where the only hardware requirement is that already provided (at relatively low-cost) by PoX architectures. 
As a result, \acro hardware cost is about 1 to 2 orders of magnitude lower than prior \CFA techniques and it is suitable for the 
low-end and ultra-low-energy MCU-s, such as MSP430 and AVR ATmega32. Furthermore, because our \acro implementation 
relies on a formally verified PoX architecture as the sole architectural component on \prv, it is also the first \CFA technique to 
offer the high-level of assurance provided by a formally verified Trusted Computing Base (TCB).

\section{Background \& Related Work}\label{sec:background}
\subsection{The Scope of Low-End Devices}\label{sec:scope}
This paper focuses on tiny CPS/IoT sensors and actuators (or hybrids thereof) with low computing power.
These are some of the smallest and weakest devices based on low-power single-core MCU-s with only a few KBytes 
of program and data memory (such as the aforementioned Atmel AVR ATmega and TI MSP430), with
$8$- and $16$-bit CPUs, typically run at $1$-$16$MHz clock frequencies, with $\approx64$ KBytes of addressable memory.
SRAM is used as data memory normally ranging in size between $4$ and $16$KBytes, while the rest of address space is 
available for program memory.  Such devices usually run software atop ``bare metal'', execute instructions in place 
(physically from program memory), and have no memory management unit (MMU) to support virtual memory.
Our implementation is based on MSP430. This choice is due to public availability of formally verified \RA~\cite{vrasedp} 
and PoX~\cite{apex} architectures implemented on OpenMSP430~\cite{OpenMSP430}, which our work relies upon. 
Nevertheless, our design rationale is applicable to other low-end MCU-s in the same class.

\subsection{Remote Attestation (\RA)}\label{sec:background_ra}
As mentioned earlier, \RA allows a trusted verifier (\vrf) to detect unauthorized code modifications (e.g., malware infections) 
on an untrusted remote device, called a prover (\dev) by remotely measuring the latter's software state.
Per Figure~\ref{fig:timeline}, \RA is typically realized as a challenge-response protocol:\\
%
	\noindent\textbf{1)-} \vrf sends an attestation request containing a challenge (\chal) to \dev. 
	This request might also contain a token derived from a secret that allows \dev to authenticate \vrf.
	
	\noindent\textbf{2)-} \dev receives the attestation request and computes an {\em authenticated integrity check} 
	over a pre-defined memory region (e.g., program memory) and \chal.
	
	\noindent\textbf{3)-} \dev returns the result to \vrf.
	
	\noindent\textbf{4)-} \vrf receives the result from \dev, and checks whether it corresponds to a valid memory state.
%
\begin{figure}[ht]
\centering
\scalebox{0.7}[0.5]{
	\fbox{
	\begin{tikzpicture}[node distance=1.5cm, >=stealth]
	\coordinate (BL)	at (0, 3);		\coordinate[left =4cm  of BL]	 (TL);
	\coordinate (Btcs)	at (0, 2.2);		\coordinate[left =4cm  of Btcs] (Ttcs);
	\coordinate (Btce)	at (0, .8);		\coordinate[left =4cm  of Btce] (Ttce);
	\coordinate (BR)	at (0, 0);		\coordinate[left =4cm  of BR]	 (TR);

	\node[above] at (BL) {\large Prover (\dev)};
	\node[above] at (TL) {\large Verifier (\vrf)};
	\coordinate (Chksum) at ($(Btcs)!0.5!(Btce)$);
	\node [right = .1cm, align=center] at (Chksum) {\small (2) Authenticated \\ \small Integrity Check};
	\coordinate (Verify)	at ($(Ttce)!0.5!(TR)$);
	\node [left =.5cm, align=center] at (Verify) {\small (4) Verify \\ \small Report};

	\draw[line width = .3cm, color=red!50]	(BL) -- (BR);
	\draw[line width = .3cm, color=blue!50] (TL) -- (TR);
	\coordinate (ReqStart) at ($(Ttcs)!0.1!(Btcs)$);
	\coordinate (ReqEnd) at ($(Ttcs)!0.9!(Btcs)$);
	\draw[thick, ->] (ReqStart) -- (ReqEnd) node [above=-.05cm, midway, sloped] {\small(1) Request};
	\coordinate (RepStart) at ($(Btce)!0.1!(Ttce)$);
	\coordinate (RepEnd) at ($(Btce)!0.9!(Ttce)$);
	\draw[thick, ->] (RepStart) -- (RepEnd) node [above=-.05cm, midway, sloped] {\small(3) Report};
\end{tikzpicture}
	}
}
	\caption{\small \RA interaction}
	\label{fig:timeline}
\end{figure}

The {\em authenticated integrity check} is usually realized as a Message Authentication Code (MAC) or a digital signature
over \dev's memory. However, these cryptographic primitives require \dev to have a unique secret key (\attkey) 
either shared with \vrf (MAC-s), or for which \vrf knows the public key (signatures). This \attkey must reside in 
secure storage, and {\bf not} be accessible to any (potentially compromised) software running on \dev, except for trusted
attestation code itself. Since most \RA threat models assume a fully compromised software state on \dev, secure storage 
implies some level of hardware support.

\RA architectures fall into three categories depending on the level of hardware support: software-based, hardware-based, and 
hybrid. Security of software-based attestation~\cite{KeJa03, SPD+04, SLS+05, SLP08} relies on strong assumptions about 
precise timing and constant communication delays, which are unrealistic in the IoT/CPS ecosystem. Hardware-based 
methods~\cite{PFM+04, TPM, KKW+12, SWP08} rely on dedicated hardware components, e.g., TPMs~\cite{TPM}, Intel SGX~\cite{sgx}, 
or ARM TrustZone~\cite{trustzone}. However, the cost of such hardware is prohibitive for low-end MCU-s. 
Hybrid \RA~\cite{smart, tytan, vrasedp} aims to achieve security equivalent to hardware-based mechanisms, 
with low(er) hardware cost. It implements the authenticated integrity ensuring function in software, while relying on minimal 
hardware support to assure that this software implementation executes properly and securely.

\subsection{Control-Flow Attestation (\CFA)}\label{sec:background_cfa}
In addition to detection of code modification via \RA, \CFA detects runtime attacks that hijack the 
program's control-flow. C-FLAT~\cite{cflat} is the earliest \CFA architecture. It uses ARM TrustZone's {\em secure 
world}~\cite{trustzone} to implement \CFA, by instrumenting the executable with context switches between TrustZone's normal 
and secure worlds. At each instruction that alters the control-flow (e.g., jump, branch, return), execution is trapped into the secure 
world and the control-flow path taken is logged into protected memory. C-FLAT targets higher-end embedded devices 
(e.g., Raspberry Pi) and its dependence on TrustZone (plus, numerous context switches) makes it unsuitable for 
low-end MCU-s targeted in this work. (Section~\ref{sec:scope} describes the scope of low-end MCU-s that we consider).

\begin{figure}[t]
	\centering
	\subfigure[Additional HW overhead (\%) in Number of Look-Up Tables]
	{\includegraphics[height=4cm,width=0.49\columnwidth]{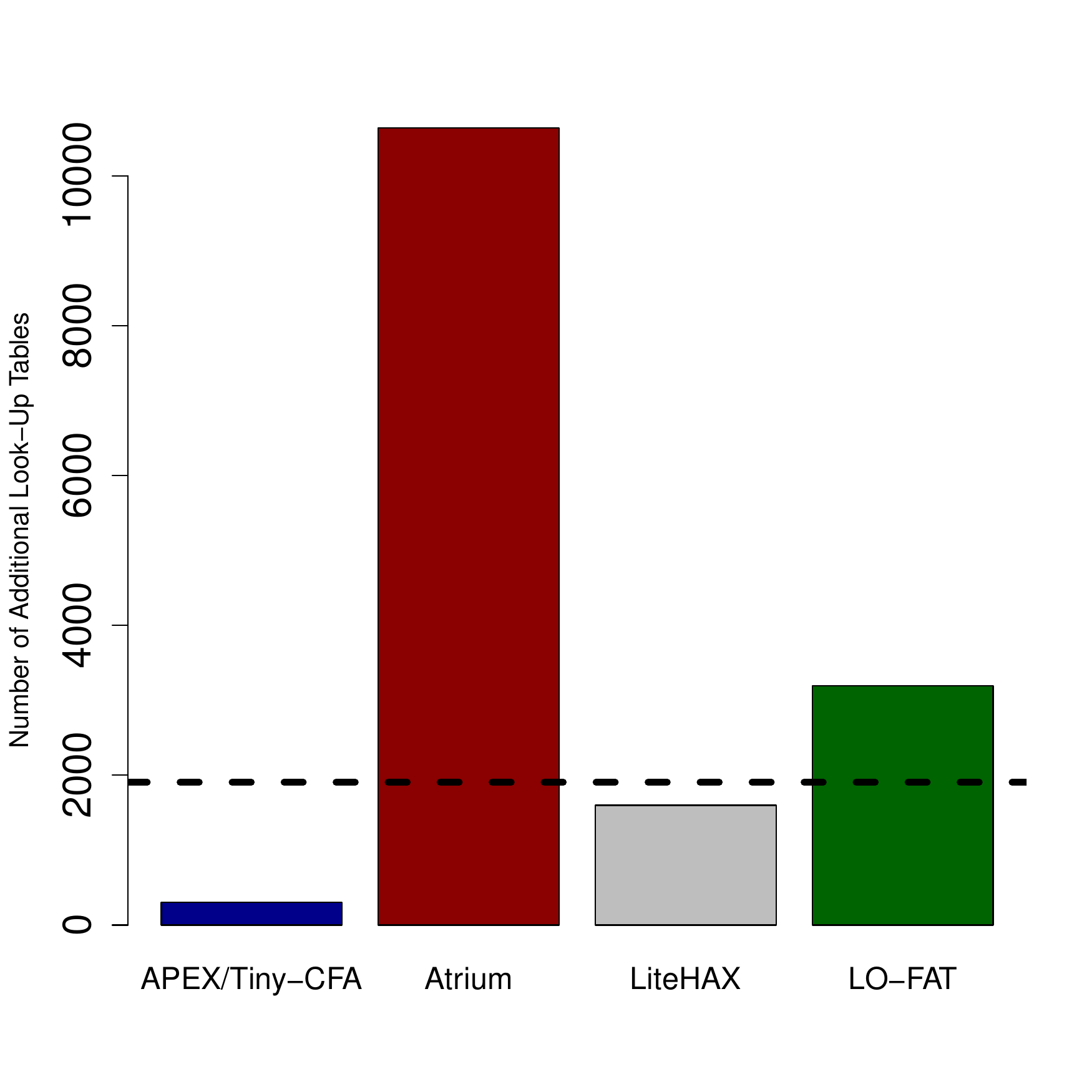}}
	\subfigure[Additional HW overhead (\%) in Number of Registers]
	{\includegraphics[height=4cm,width=0.49\columnwidth]{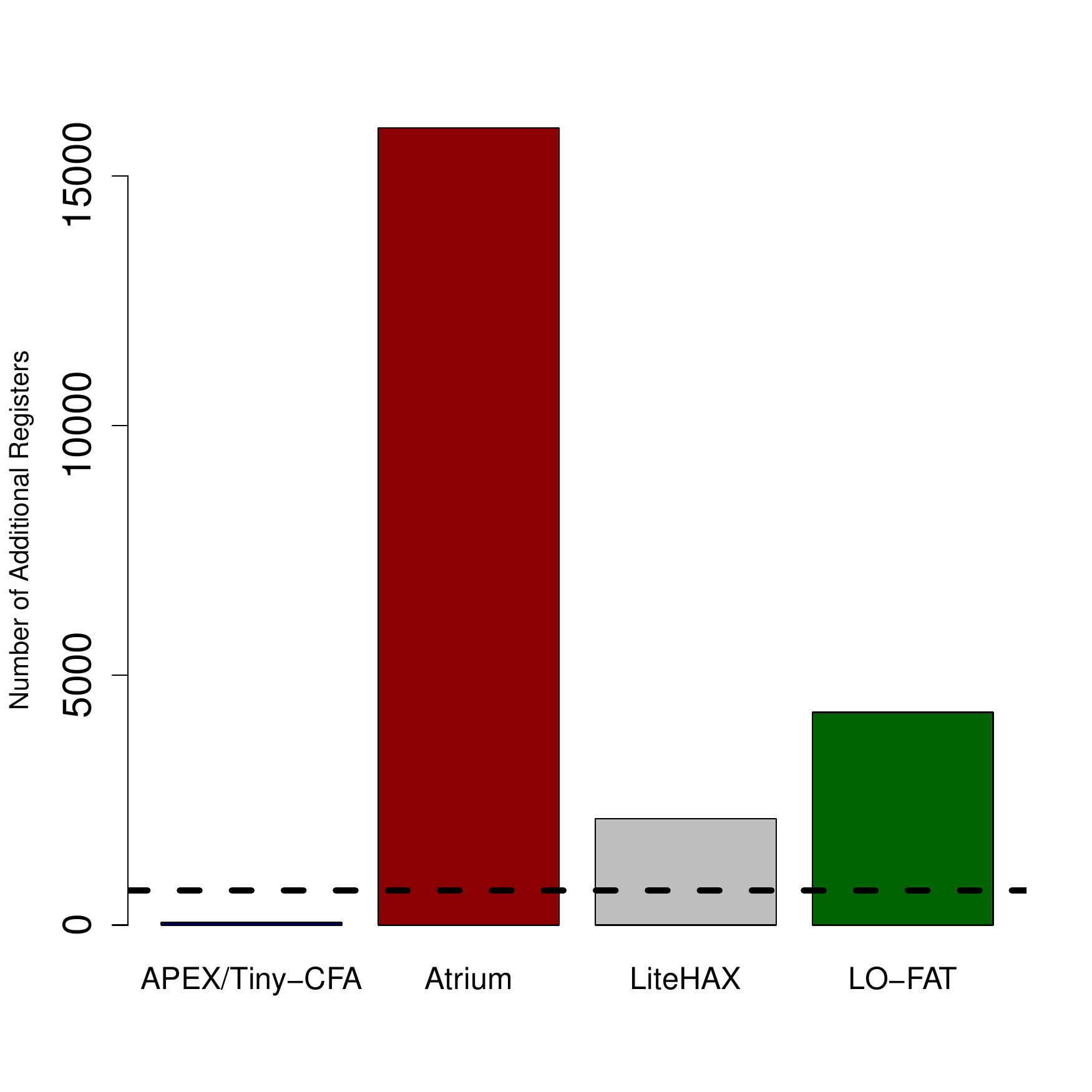}}
	\caption{Overhead comparison between \CFA architectures and PoX (APEX). Dashed lines represent the total hardware cost of MSP430 core itself. Hardware costs are as reported in the original papers~\cite{dessouky2017fat,dessouky2018litehax,zeitouni2017atrium,apex}.}\label{fig:comparison}
\end{figure}

To remove the TrustZone dependence, subsequent \CFA techniques: LO-FAT~\cite{dessouky2017fat} and LiteHAX~\cite{dessouky2018litehax},
implement \CFA using stand-alone hardware modules: a branch monitor and a hash engine. Atrium~\cite{zeitouni2017atrium} 
enhances aforementioned \CFA techniques, securing them against physical adversaries that intercept instructions as they are fetched 
to the CPU. Though less expensive than C-FLAT, such hardware components are still not viable for low-end MCU-s, since their cost 
(in terms of price, size, and energy consumption) is typically higher than the cost of a low-end MCU itself. 
This is evident from Figure~\ref{fig:comparison}, which compares hardware costs -- in terms of Look-Up Tables (LUTs) and numbers of 
Registers -- of aforementioned \CFA techniques and the total hardware cost of the OpenMSP430's core itself, represented by 
dashed lines.

\subsection{Proofs of Execution (PoX)}\label{sec:background_pox}
PoX augments \RA capability by proving to \vrf that: (1) the expected executable is stored in program memory,
and (2) this code indeed executed, and any claimed outputs were produced by its timely and authentic execution.

The first PoX architecture targeting low-end MCU-s was recently proposed in APEX~\cite{apex}.
APEX implements a hardware module controlling the value of a $1$-bit flag called $EXEC$, that cannot be written by any software. 
A value $EXEC=1$ indicates to \vrf that attested code \emph{must} have executed successfully, between the time when the 
challenge was received from \vrf (recall the \RA protocol from Section~\ref{sec:background_ra}) and the time when the 
\RA measurement occurs (via authenticated integrity ensuring function). Similarly, when it receives an attestation reply with $EXEC=0$, 
\vrf can conclude that execution of said code did not occur, or that execution (or its output) was tampered with.
In APEX, the \RA measurement covers: {\bf(i)} the $EXEC$ flag itself; {\bf(ii)} the region where this execution's output is saved 
(output region -- $OR$); and {\bf(iii)} the executable itself (stored in the executable region -- $ER$). Thus, security of the 
underlying \RA architecture guarantees that the contents of these memory regions cannot be forged/spoofed to something 
different from their values at time of the attestation computation. In turn, APEX considers a code to execute properly (and sets 
$EXEC=1$) if and only if:

	\noindent\textbf{1)-} Execution is atomic (i.e., uninterrupted), from the executable's first instruction (legal entry $ER_{min}$), 
	to its last instruction (legal exit $ER_{max}$);
	
	\noindent\textbf{2)-} Neither the executable ($ER$), nor its produced outputs $OR$ are modified in 
	between the execution and subsequent \RA computation;
	
	\noindent\textbf{3)-} During execution, data-memory (including $OR$) cannot be modified, by means other than the 
	executable in ER itself, e.g., no modifications by other software or Direct Memory Access controllers.

These conditions mean that $EXEC=1$ assures that memory contents (of $ER$ and $OR$) are consistent between $ER$'s 
code execution and respective attestation, and that execution itself is untampered, e.g. via interruptions, or modification of 
intermediate results in data memory. $ER$ and $OR$ locations and sizes are configurable, allowing for PoX of arbitrary code 
and output sizes. APEX implementation is built atop the formally verified hybrid \RA architecture \vrased~\cite{vrasedp}, 
and APEX hardware module is itself formally verified to adhere to a set of formal logic specifications. These properties, 
along with \vrased verified guarantees, are proven sufficient to imply a security definition (stated using the cryptographic 
security game framework~\cite{crypto_book}) for unforgeable of proofs of execution. Due to space constraints, we do not 
overview APEX proofs and refer the interested reader to~\cite{apex}.

As discussed in~\cite{apex}, similar to Trusted Execution Environments (TEEs) targeting higher-end platforms (e.g., 
Intel SGX~\cite{sgx} and ARM TrustZone\cite{trustzone}), APEX assumes executable correctness, i.e., the user is 
responsible for programming \prv with bug-free and memory-safe code. Hence, by default, APEX does not provide 
any security against runtime (aka control-flow) attacks. In this work, we bridge this gap by introducing an automated code 
instrumentation technique that leverages APEX to implement \CFA in low-end MCU-s. In other words, we show that 
\CFA on top of APEX (or more generally any PoX), without any additional hardware requirement, is both possible and affordable.
As a clear advantage over prior techniques, our approach requires $5.4$ times fewer additional LUTs and $50$ times 
fewer additional registers than the second cheapest approach -- LiteHAX; see comparison of APEX hardware overhead with 
other \CFA techniques in Figure~\ref{fig:comparison}).

\section{\acro}
\acro couples a formally verified PoX with code instrumentation to obtain \CFA. It uses APEX PoX that ties the 
executed code to its output, stored in a data-memory range of configurable size, called $OR$.
The basic idea is to instrument the code to produce a log of the program control-flow path, and make it a part of 
output. The program instrumentation writes the destination address of each control-flow altering instruction 
into $OR$. We denote this control-flow log as \cflog.

\begin{figure}[t]
	\centering
    \includegraphics[height=3cm,width=0.49\columnwidth]{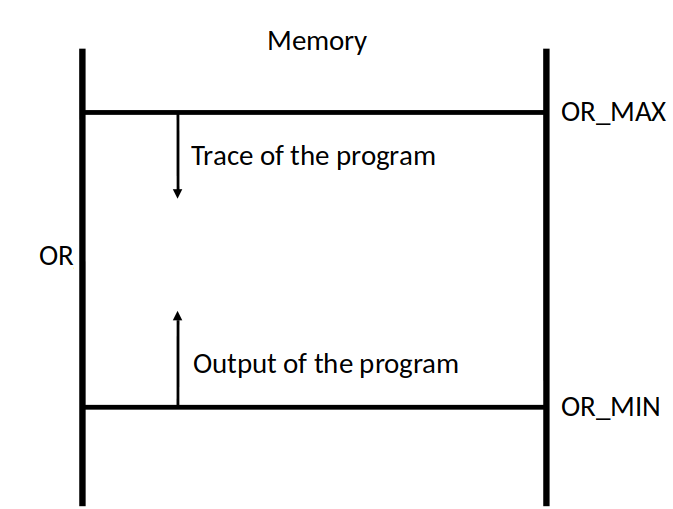}
	\caption{OR region used to store regular program outputs and \cflog.}\label{fig:OR_CFA}
\end{figure}

As shown in Figure~\ref{fig:OR_CFA}, in \acro, both regular program outputs and \cflog are written to $OR$.
Recall from Section~\ref{sec:background_pox} that $OR$ size/location is configurable. Hence, \vrf can chose $OR$ to be large 
enough to fit both the regular program output and its expected \cflog. Note that, in any \CFA scheme, \vrf must have 
\textit{a priori} knowledge of the expected/benign control-flows and their sizes. Therefore, the appropriate $OR$ size is 
trivially obtained by adding the regular output and \cflog sizes. The regular program output is written to $OR$ normally,
bottom-to-top of $OR$, as in APEX. Whereas, \acro instrumentation writes \cflog to $OR$ from top to bottom. This strategy is 
similar to how stack and heap are handled in RAM and it assures that the program output and \cflog do not interfere or 
overlap with each other, as long as $OR$ is appropriately sized.

We believe that this general idea is both intuitive and sensible; it guides \acro's design.
However, ensuring that \acro results in a {\em secure} \CFA design is significantly more challenging.
To see why, note that \textbf{the executable to be attested}, (i.e., security-critical code stored in $ER$) 
{\bf is itself subject to control-flow attacks}. Thus, any values logged to \cflog by the instrumented executable 
can, in principle, be modified as part of a control-flow attack. In other words, \acro's approach is only secure is 
\cflog is an \textbf{append-only log}. Otherwise, upon completion of its nefarious tasks, a control-flow attack can 
overwrite \cflog to reflect a benign or expected control-flow, erasing any trace of the compromised control-flow and thus
fool \vrf. In higher-end \CFA architectures (e.g., C-FLAT~\cite{cflat}), this property is obtained by logging the control-flow 
to dedicated secure memory, which is never accessible to untrusted/application code, e.g., C-FLAT uses TrustZone's secure world. 
However, as discussed in Sections~\ref{sec:intro} and~\ref{sec:background}, low-end MCU-s cannot afford such expensive 
security features. Below, we detail how \acro can be made secure by relying exclusively on PoX and instrumentation, 
thus retaining its suitability for low-end MCU-s.

\subsection{Design Rationale \& Security}\label{sec:design}
We now discuss \acro design rationale and security  properties (\textbf{P1-P6}) at high-level. 
Implementation details of an instance of \acro on MSP430 are further specified in Section~\ref{sec:implementation}.
We postulate the properties that ensure that control-flow attacks are always detected under the following comprehensive adversarial model:

\emph{\textbf{Adversarial Model --} we assume that the adversary controls \dev's entire software state, including code 
and data. \adv\ can modify any writable memory and read any memory that is not explicitly protected by hardware-enforced 
access control rules (e.g., APEX rules). Program memory modifications can be performed to change instructions, 
while data memory modifications may trigger control-flow attacks. Adversarial modifications are allowed 
before, during, or after the execution of the program whose control-flow is to be attested.}

\textbf{[P1]: Integrity of Code, Instrumentation and Output --}  Clearly, any instrumentation-based approach is only sound if 
unauthorized modifications to the instrumented code itself (e.g., to remove instrumentation) are detectable. Detection of 
modifications is offered by the underlying \RA and PoX architectures (see Section~\ref{sec:background}). In particular, 
these architectures guarantee that any unauthorized code modification is detected by \vrf. They also guarantee that 
modifications to attested executable's output ($OR$ -- which includes \cflog) are only possible if done by the attested 
executable itself, during its execution.

\textbf{[P2]: Secure logging of control-flow instructions--}
The first step in \acro, is to instrument all control-flow altering instructions to log their destinations to \cflog, in OR.
\cflog is implemented as a stack, from the highest value in OR ($OR_{max}$) growing downwards, as shown in Figure~\ref{fig:OR_CFA}.
The pointer to the top of this stack is stored in a dedicated register \rtop. Each control-flow instruction is then instrumented with 
additional instructions to push its destination address to this stack, i.e.:  (i) write the destination of address to the memory location 
pointed to by \rtop; and \textbf{(ii)} decrement \rtop. At instrumentation time, the assembly code of the executable is inspected 
to assure that no other instructions utilize the MCU register \rtop. In all practical examples considered in this work, executables 
have at least one free register available. If no such register exists by default, the code can be recompiled to free up one register.

\textbf{[P3]: Secure logging of conditional branches --}
Conditional branches determine control-flow at runtime, depending on a result of a conditional statement, e.g., a comparison or 
equality test. These instructions are used to implement \texttt{loops} and \texttt{if-then-else} statements used in high-level 
languages. Conditional branches are instrumented by pushing to \cflog's stack (using the same method as in {\bf P2}) 
the possible destinations as well as the result of the conditional statement. This way, by inspecting \cflog, \vrf can determine the 
exact path taken by the conditional branch.

\textbf{[P4]: Write safety --} 
Write operations are dangerous since they can be used during an attack to overwrite \cflog, thus
hiding the compromised control-flow from \vrf. Direct writes (which modify constant addresses) are easy to deal with, because 
they can be statically inspected for safety at instrumentation time. In particular, the instrumenter can verify that no direct writes 
modify \cflog reserved addresses in $OR$. Indirect writes modify memory addresses determined at runtime.
Consequently, they require instrumentation to check their safety, also at runtime. After each indirect write, \acro instrumentation  
introduces instructions to check whether the write destination is within \cflog by checking if the write destination is within the 
range $[\rtop,OR_{max}]$ -- the memory range currently in use to store \cflog. Upon detection of an illegal write, execution is 
halted, implying an invalid control-flow.

\textbf{[P5]: Wrap-around attack protection --} 
Given the inability to modify \cflog due to checks performed in previous steps, the last resort for a control-flow attack to go 
undetected is to keep executing control-flow instructions until \rtop value overflows and wraps-around, coming back to its initial 
value $\rtop = OR_{max}$ and overwriting of \cflog. To protect against such attacks, modifications to \rtop have an additional check, 
ensuring that whenever \rtop points to an instruction outside $OR$ range, execution is halted.

\textbf{[P6]: \rtop initialization verification --}
Previous properties rely on the fact that \rtop is initialized as $\rtop = OR_{max}$ at the start of execution, to assure that \cflog is 
indeed stored in $OR$. However, performing this initialization inside the executable being attested allows for control-flow attacks 
that jump back to the \rtop initialization code to reset \rtop in the middle of the execution. Instead of initializing \rtop inside the 
attested executable, \acro instruments the executable to check if \rtop has been previously properly initialized to $\rtop = OR_{max}$. 
In turn, the caller application becomes responsible for initializing \rtop appropriately, making control-flow attacks that re-initialize 
\rtop to reset \cflog impossible, sine they require jumping outside of the executable range -- $ER$ -- which is detected by APEX
as a violation.

\emph{\textbf{\underline{Security Argument}:} 
Let \program denote a procedure/function/code-segment for which execution and control-flow need to be attested. 
Properties {\bf P2 \& P3} assure that all changes to the control-flow of \program are logged to \cflog at runtime. 
Then, by inspecting an authentic (untampered) \cflog, \vrf can determine the exact control-flow taken by that particular \program
execution. Meanwhile, properties {\bf P5 \& P6} guarantee that \cflog is stored inside $OR$, within $[\rtop, OR_{max}]$ range. 
Property {\bf P4} detects any illegal writes during execution that attempt to modify \cflog, i.e., writes to $[\rtop, OR_{max}]$ range. 
Hence, for a given execution of \program, the combination of {P4, P5 \& P6} guarantees that each written value can never be 
overwritten or deleted from \cflog. Finally, {\bf P1}, inherited from the underlying PoX architecture, assures that neither \program
instructions (including instrumentation), nor its output (including \cflog) can be modified by other means (e.g., other software on \prv, 
interrupts, DMA) before, during, or after execution. Any such attempt is detectable by \vrf, because it causes APEX to set 
$EXEC=0$; recall the $EXEC$ flag behavior described in Section~\ref{sec:background_pox}. Therefore, \acro properties 
{\bf P1-P6} suffice to implement secure \CFA, under the aforementioned adversarial model. \qed
}

\subsection{Optimizations}
In practice, \cflog size determines the practicality of \acro due to the resource-constrained nature of low-end MCU-s, especially, 
with respect to memory size. In fact, although secure, the approach described thus far tends to bloat rapidly for control-flow
intensive code segments, e.g., loops with many iterations. In this section, we discuss two simple optimizations \textbf{(O1 \& O2)}
that significantly reduce \cflog size without sacrificing overall security.

\textbf{O1- Static Control-Flow Instructions --}
We observe that control-flow instructions with constant destination addresses (determined statically in the code) need not 
be logged to \cflog, as their effect on the program control-flow can not change at runtime. This removes the need to log operations,
such as usual function calls (with exception of callbacks), fixed-address \texttt{go-to-s}, and similar.

\textbf{O2- Loops --}
Loops are challenging to log efficiently due to their high number of control-flow operations. For instance, consider a delay function 
based on \textit{busy-wait}, commonly used in MCU code. It essentially consists of a loop that increments a counter up to a 
certain constant. The higher the delay, the higher the number of iterations, implying the higher the number of control-flow instructions 
to be logged. In turn, even a simple loop, such as a 1-second delay, would require 
millions of iterations (assuming typical clock frequencies on the order of MHz) resulting in millions of symbols logged to \cflog.
To deal with such cases, we introduce an optimization that removes the requirement to store each control-flow instruction for 
loops for which number of iterations can be predicted statically, at instrumentation time.

Specifically, \acro instrumenter inspects each conditional branch. For each loop branch instruction instruction $bi$, changing the 
control-flow to destination instruction $di$, the instrumenter inspects all instructions in the range $[bi,di]$. If no indirect control-flow 
instructions exist in this range, the number of iterations caused by such a loop can be determined exclusively by checking the 
branch condition and the variables involved in this condition. Therefore, instead of logging
each branch at every iteration, \acro simply logs the condition itself, only once. This allows \vrf to learn the exact control-flow 
generated by a loop (i.e., \# iterations) without bloating \cflog. In our 1-second delay example, instead of logging millions of 
symbols, the loop would log just a couple of bytes, corresponding to the loop exit condition (typically, a comparison to a constant, 
e.g., $i < 10^6$). This optimization also applies to loops used in common memory/array manipulations, e.g., 
in \texttt{memset}, and \texttt{memcpy}.

\subsection{Implementing \acro}\label{sec:implementation}
We now describe how properties {\bf P1-P6} are securely implemented via automatic assembly instrumentation on the MSP430 MCU.
Our instrumenter is implemented in \texttt{Python} with approximately $300$ lines of code.

Figure~\ref{fig:return_instrumentation} shows the instrumentation of indirect control-flow instructions: \textit{return} in this particular example.
It writes the return address, which in MSP430 assembly must be loaded to register $r1$ before \texttt{ret} is called, to \cflog.
In our implementation $\rtop=r4$. Hence, the content of $r1$ (destination address) is copied to the address 
pointed to by \rtop in $OR$, as required by {\bf P2}. To also enforce {\bf P5}, upon writing to the address of \rtop, and moving \rtop 
to point to the next address, the comparison at line 3 checks if \rtop is still inside $OR$,
otherwise exiting the program, by jumping to an exit instruction at line 4. 

Figure~\ref{fig:making_writes_great_again} depicts the instrumentation of indirect write instructions to enforce {\bf P4}. 
Upon writing to a given memory location (address pointed to by $r14$, in this example), checks are performed to determine if this write 
operation did not modify \cflog memory range: $[\rtop, OR_{max}]$. If an illegal write occurs, program execution is halted 
(at line 5) and a control-flow attack attempt is detected.

Figure~\ref{fig:initial_check} shows the instrumentation, required by {\bf P6}, at the beginning of the code segment. 
It ensures that \rtop is properly initialized, otherwise halting execution at line 3.

Finally, Figure~\ref{fig:conditional_branch} depicts the instrumentation required by {\bf P3}. It logs to \cflog the results of 
conditional statements. Note that, after a conditional statement (e.g., at line 1) evaluation, the result is stored in the status register $r2$. Hence, the content of $r2$ is written to \cflog (line 2), since it is sufficient to determine the destination 
of the conditional branch. The same check to enforce {\bf P5} in Figure~\ref{fig:return_instrumentation}, is also performed in 
this case, because information is being written to \cflog. Since this check itself overwrites $r2$, the original value of $r2$ 
needs to be retrieved (at line 6) before the actual branch instruction at line 7.

\emph{\textbf{Remark:} \acro can not be abused by control-flow attacks that jump in the middle of the instrumentation instructions. 
Such an illegal jump is logged to \cflog and is thus detectable by \vrf. Since \rtop never 
%
%
retracts (within a given execution), write checks (see Figure~\ref{fig:making_writes_great_again}) make it impossible to delete any information 
logged to \cflog, including jumps into the middle of instrumented code instructions.}

\lstdefinelanguage
   [x64]{Assembler}     
   [x86masm]{Assembler} 
   %
   {morekeywords={mov.b,jn,jlo,cmp.b}}
   
\lstset{language=[x64]{Assembler},
	basicstyle={\tiny\ttfamily},
	showstringspaces=false,
	frame=single,
	xleftmargin=2em,
	framexleftmargin=3em,
	numbers=left, 
	numberstyle=\tiny,
	commentstyle={\tiny\itshape},
	keywordstyle={\tiny\ttfamily\bfseries},
	keywordstyle=\color{blue}\tiny\ttfamily\ttfamily,
	stringstyle=\color{red}\tiny\ttfamily,
        commentstyle=\color{black}\tiny\ttfamily,
        morecomment=[l][\color{magenta}]{\%},
        breaklines=true
}

\begin{figure}
\begin{minipage}{0.49\linewidth}
\begin{lstlisting}[xleftmargin=.13\textwidth, xrightmargin=.13\textwidth]




ret
\end{lstlisting}
\centering
\scriptsize{(a) Original}
\end{minipage}
\begin{minipage}{0.49\linewidth}
\begin{lstlisting}[xleftmargin=.13\textwidth, xrightmargin=.13\textwidth]
mov r1, @r4
dec r4
cmp #OR_MIN, r4
jn .L11
ret
\end{lstlisting}
\centering
\scriptsize{(b) Instrumented}
\end{minipage}
\caption{Instrumentation example: indirect control-flow instructions.}\label{fig:return_instrumentation}
\end{figure}

\begin{figure}
\begin{minipage}{0.49\linewidth}
\begin{lstlisting}[xleftmargin=.13\textwidth, xrightmargin=.13\textwidth]
mov.b r15, @r14





...
\end{lstlisting}
\centering
\scriptsize{(a) Original}
\end{minipage}
\begin{minipage}{0.49\linewidth}
\begin{lstlisting}[xleftmargin=.13\textwidth, xrightmargin=.13\textwidth]
  mov.b r15, @r14
  cmp r4, r14
  jlo .L12
  cmp #OR_MAX, r14
  jlo .L11
.L12:
...
\end{lstlisting}
\centering
\scriptsize{(b) Instrumented}
\end{minipage}
\caption{Instrumentation example: indirect write instructions.}\label{fig:making_writes_great_again}
\end{figure}

\begin{figure}
\begin{minipage}{0.49\linewidth}
\begin{lstlisting}[xleftmargin=.13\textwidth, xrightmargin=.13\textwidth]
application:


...
\end{lstlisting}
\centering
\scriptsize{(a) Original}
\end{minipage}
\begin{minipage}{0.49\linewidth}
\begin{lstlisting}[xleftmargin=.13\textwidth, xrightmargin=.13\textwidth]
application:
  cmp #OR_MAX, r4
  jne .L11
...
\end{lstlisting}
\centering
\scriptsize{(b) Instrumented}
\end{minipage}
\caption{Instrumentation example: \rtop initialization check.}\label{fig:initial_check}
\end{figure}

\begin{figure}[hbtp]
\begin{minipage}{0.49\linewidth}
\begin{lstlisting}[xleftmargin=.13\textwidth, xrightmargin=.13\textwidth]
  cmp.b #64, r15


  
  
  
  jne .L2
...
\end{lstlisting}
\centering
\scriptsize{(a) Original}
\end{minipage}
\begin{minipage}{0.49\linewidth}
\begin{lstlisting}[xleftmargin=.13\textwidth, xrightmargin=.13\textwidth]
  cmp.b #64, r15
  mov r2, @r4
  dec r4
  cmp #OR_MIN r4
  jn .L11
  mov l(r4), r2 
  jne .L2
...
\end{lstlisting}
\centering
\scriptsize{(b) Instrumented}
\end{minipage}
\caption{Instrumentation example: conditional branches.}\label{fig:conditional_branch}
\end{figure}

\section{Case Study \& Evaluation}\label{sec:eval}
\subsection{Case Study: Control-Flow Attacks in Low-End MCU-s}\label{sec:case_studies}
Control-flow attacks can be extremely harmful, especially, for low-end devices used for safety-critical tasks.
To illustrate this point, we show an attack on a medical syringe pump application implemented on a low-end MCU.
For clarity, we focus on a simplified version of the OpenSyringePump 
application\footnote{Available at: \url{https://github.com/naroom/OpenSyringePump}}. 
Later, in Section~\ref{sec:eval}, we evaluate \acro on three applications, including the original 
OpenSyringePump code, which is longer and more complex than the example used here. 
OpenSyringePump was also used to motivate and evaluate prior \CFA approaches, e.g., C-FLAT.

Consider the \texttt{C} code segment in Figure~\ref{fig:attack_example}. In this application, the MCU is connected through the 
general-purpose input/output (GPIO) port $P3OUT$ (used at lines $5$ and $8$) to an actuator, responsible for injecting a 
given dose of medicine, determined in software, according to commands received through the network, e.g., from a remote physician.
The function \texttt{injectMedicine} injects the appropriate dosage according to the variable \texttt{dose}, by triggering actuation 
for an amount of time corresponding to the value stored in \texttt{dose}. To guarantee a safe dosage, the \texttt{if} statement 
(at line 4) assures that the maximum injected dosage is $9$, thus preventing overdosing.

Dosage is determined according to a list of values, e.g., symptom severity measures received from a remote physician. 
The function \texttt{parseCommands} (line 11) is responsible for making a copy of the received values and processing 
them to determine appropriate dosage. However, this function can also be used to trigger a buffer overflow attack, 
leading to a malicious control-flow path.
Specifically, because the size of \texttt{copy\_of\_commands} is static and equal to $5$, an input array of larger size 
can cause other values in the program's stack to be overwritten, beyond the area allocated for \texttt{copy\_of\_commands}, 
and including the memory location storing the return address of \texttt{parseCommands}. In particular, the return address is 
overwritten with the value of \texttt{recv\_commands[5]}. By setting the content of \texttt{parseCommands[5]} to the address of 
line 5 in Figure~\ref{fig:attack_example}, such an attack causes the control-flow to jump directly to line 5 
(when \texttt{parseCommands} returns), skipping the safety check at line 4, and potentially overdosing the patient.

The above attack example is detectable neither by static \RA techniques nor by PoX techniques, since expected (unmodified) 
code still executes in its entirety, yet in an unexpected order. \acro, however, detects such control-flow attacks, because  
the instrumentation of indirect control-flow instructions (e.g., \texttt{return} in Figure~\ref{fig:return_instrumentation}) commits the 
maliciously overwritten return address to \cflog.

In Section~\ref{sec:eval} we evaluate \acro performance in 3 realistic safety-critical applications: (1) OpenSyringePump -- 
the full implementation of our toy example in Figure~\ref{fig:attack_example}; (2) 
FireSensor~\footnote{Available at: \url{https://github.com/Seeed-Studio/LaunchPad_Kit/tree/master/Grove_Modules/temp_humi_sensor}} -- 
a fire detector based on temperature and humidity sensors; and (3) 
UltrasonicRanger~\footnote{Available at: \url{https://github.com/Seeed-Studio/LaunchPad_Kit/tree/master/Grove_Modules/ultrasonic_ranger}} -- 
a sensor used by parking assistants for obstacle proximity measurement.

\lstset{language=C,
	basicstyle={\tiny\ttfamily},
	showstringspaces=false,
	frame=single,
	xleftmargin=2em,
	framexleftmargin=3em,
	numbers=left, 
	numberstyle=\tiny,
	commentstyle={\tiny\itshape},
	keywordstyle={\tiny\ttfamily\bfseries},
	keywordstyle=\color{blue}\tiny\ttfamily\ttfamily,
	stringstyle=\color{red}\tiny\ttfamily,
        commentstyle=\color{black}\tiny\ttfamily,
        morecomment=[l][\color{magenta}]{\%},
        breaklines=true
}
\vspace*{-0.3cm}
\begin{figure}[hbtp]
\centering
\begin{minipage}{0.94\linewidth}
\begin{lstlisting}[]
int dose = 0;

void injectMedicine(){
  if (dose < 10){  //safety check preventing overdose
    P3OUT = 0X1;	
    delay(dose*time_per_dose_unit);
  }
  P3OUT = 0x0;
}

void parseCommands(int *recv_commands, int lenght){
  int copy_of_commands[5];
  memcpy(copy_of_commands, recv_commands, lenght);
  dose = processCommands(copy_of_commands);
  return;
}
\end{lstlisting}
\end{minipage}
\vspace{-2mm}
\caption{Safety critical application exploitable by control-flow attacks.}\label{fig:attack_example}
\end{figure}
\vspace*{-0.3cm}
\subsection{Experimental Results}\label{sec:runtime}
Recall that, since \acro requires no hardware support beyond that already provided by APEX~\cite{apex}, its 
hardware costs remain consistent with Figure~\ref{fig:comparison}. Therefore, this section focuses on other costs: 
code size increase, runtime overhead, and \cflog size. As mentioned in Section~\ref{sec:case_studies}, our evaluation 
instantiates \acro on MSP430 with three real-world, publicly available, and safety-critical use cases: \texttt{SyringePump}, \texttt{FireSensor}, and 
\texttt{UltrasonicRanger}. Tables~\ref{table:original} and~\ref{table:instrumented} present experimental results for these three 
applications in their unmodified forms and when instrumented by \acro.
In each case, the attested execution corresponds to one iteration of the application's main loop (i.e., the application can report to \vrf with the attestation response once per iteration), involving the respective 
sensing and actuation tasks.
\begin{table}[hbtp]
\footnotesize
\begin{tabular}{|l|l|l|l|}
\hline
          & SyringePump & FireSensor & UltrasonicRanger \\ \hline
{\bf Code Size} & 218 bytes & 434 bytes      &  238 bytes           \\ \hline
{\bf Runtime}   &   159644 cycles & 20919 cycles &  2799 cycles     \\ \hline
\end{tabular}
\vspace{0.1cm}
\caption{{\footnotesize Original application costs}}\label{table:original}
\end{table}
\vspace{-0.7cm}
\begin{table}[hbtp]
\footnotesize
\begin{tabular}{|l|l|l|l|}
\hline
          & SyringePump & FireSensor & UltrasonicRanger \\ \hline
{\bf Code Size} & 416 bytes & 790 bytes & 442 bytes \\ \hline
{\bf Runtime}   &  162218 cycles  & 31818 cycles & 3027 cycles \\ \hline
{\bf \cflog size}   &  400 bytes   & 2068 bytes & 30 bytes \\ \hline
\end{tabular}
\vspace{0.1cm}
\caption{\footnotesize Instrumented application costs}\label{table:instrumented}
\end{table}
\vspace*{-0.4cm}

In all three cases, code size increases by $\approx 80\%$, while \cflog size ranges between $30$ and 
$2k$ Bytes, and runtime overhead varies between $\approx 2\%$ and $\approx 50\%$.
CF-Log size depends on the number of control-flow transfers occurring in the application. 
Programs performing simple tasks need smaller log size ($< 1 k$ bytes ), 
while those with complex tasks would need larger log sizes.

\acro exhibits lower runtime overhead than C-FLAT~\cite{cflat}. C-FLAT is only evaluated using the \texttt{SyringePump} example, 
and its reported runtime overhead is $\approx 76\%$, due to instrumentation of trampolines and context switches; see~\cite{cflat} for details. 
Meanwhile, in all considered applications, \acro runtime overhead remains below $\approx 50\%$. This is justified by: (1) simpler design 
that does not rely on trampoline hypercalls or context switches, and (2) optimization {\bf O2}, which removes per-iteration instrumentation 
away from delay loops. Since delay loops are used frequently in sensing/actuation applications, this optimization comes in handy in 
most practical scenarios. However, we do not compare runtime overhead of \acro with Lo-FAT and LiteHAX since these two techniques 
do not instrument code, instead detecting branches in hardware.

In summary, experimental results indicate that, in all sample applications, instrumented executables remain well within the 
capabilities of low-end MCU-s, thus supporting \acro's practicality.

\vspace{-0.1cm}
\section{Conclusions}\label{sec:conclusion}
\vspace{-0.1cm}
We designed, implemented and evaluated \acro: a low-cost \CFA approach targeting low-end MCU-s. 
\acro couples a formally verified PoX architecture with automated code instrumentation to yield an effective low-cost
\CFA. We argued security of \acro and demonstrated, via a MSP430-based implementation, its ability to detect control-flow 
attacks.

\noindent {\bf Acknowledgments:} We thank DATE'21 anonymous referees for their helpful comments.
This research was supported in part by funding from Army Research Office (ARO), under contract
W911NF-16-1-0536 and Semiconductor Research Corporation (SRC), under contract 2019-TS-2907.

\vspace{-2mm}
\bstctlcite{IEEEexample:BSTcontrol}
\bibliographystyle{IEEEtranS}

{\small
\linespread{0.89}
\bibliography{IEEEabrv,references}
}
\end{document}